\documentclass[aps,pra,twocolumn,groupedaddress]{revtex4-1}
\usepackage{graphicx}
\usepackage{dcolumn}
\usepackage{bm}
\usepackage{amsthm}
\usepackage{multirow}

\begin{document}

\title{A Low-Overhead Hybrid Approach for Universal Fault-Tolerant Quantum Computation}


\author{Eesa Nikahd}
\email{nikahd@aut.ac.ir}
\author{Morteza Saheb Zamani}
\email{szamani@aut.ac.ir}
\author{Mehdi Sedighi}
\email{msedighi@aut.ac.ir}

\affiliation{Quantum Design Automation Lab \linebreak Amirkabir University of Technology, Tehran, Iran}


\date{\today}


\begin{abstract}
As there is no quantum error correction code with universal set of transversal gates, several approaches have been proposed which, in combination of transversal gates, make universal fault-tolerant quantum computation possible. Magic state distillation, code switching, code concatenation and pieceable fault-tolerance are well-known examples of such approaches. However, the overhead of these approaches is one of the main bottlenecks for large-scale quantum computation. In this paper, a hybrid approach is proposed which combines the code concatenation technique with the other mentioned approaches. The proposed approach outperforms code concatenation in terms of both number of qubits and error threshold and also significantly reduces the resource overhead of code switching, magic state distillation and pieceable fault-tolerance at the cost of reducing the effective distance of the concatenated code for implementing non-transversal gates.
\end{abstract}


\maketitle


\section{\label{sec:intro}Introduction}

Quantum computers have the potential to efficiently solve certain problems such as integer factorization \cite{shor1994algorithms1} and simulation of quantum systems \cite{zalka1998efficient2} which are prohibitively time-consuming using classical computers. This computational advantage of quantum systems comes from the unique quantum mechanical properties such as superposition and entanglement, which have no classical analogue \cite{nielsen2010quantum3}.

Quantum bits or qubits are the fundamental units of information in quantum computing. Unfortunately, qubits are fragile and tend to lose their information due to the environmental noise resulting in decoherence \cite{nielsen2010quantum3}\cite{unruh1995maintaining4}. Furthermore, the physical implementations of quantum operations in any technology are imperfect \cite{unruh1995maintaining4}\cite{mazzola2010sudden5}. Quantum noise, due to decoherency of quantum states and imperfect quantum operations, is the most important challenge in constructing large-scale quantum computers \cite{nielsen2010quantum3}\cite{metodi2006quantum6}\cite{ahsan2015architecture7}.

The most common approach to cope with this challenge is the use of quantum error correction codes and fault-tolerant operations to perform quantum computation. In this approach, a logical qubit is encoded into multiple physical qubits using a suitable error correction code. Logical operations are applied directly on the encoded qubits in such a manner that decoding is not required. After that, if a qubit becomes erroneous, that qubit can be corrected using repeated application of the quantum error correction procedure. The logical operations can potentially spread errors due to the interactions among qubits. Therefore, to preserve the veracity of computation, these operations must be implemented fault-tolerantly in such a way that they do not propagate errors from a corrupted qubit to multiple qubits in a codeword.

Transversal implementation of logical gates is widely considered as a simple and efficient method for fault-tolerant quantum computation (FTQC) \cite{shor1996fault9}\cite{anderson2014fault10}, where a transversal gate refers to a gate which does not couple qubits inside a codeword. Unfortunately, there is no quantum code with a universal set of transversal gates \cite{eastin2009restrictions11}. Several approaches have been proposed which, in combination with transversal gates, make FTQC possible. Magic state distillation \cite{bravyi2005universal12}, gauge fixing \cite{paetznick2013universal17}\cite{bombin2015gauge18}, code switching \cite{anderson2014fault10}\cite{stephens2008asymmetric13}\cite{choi2015dual14}, code concatenation \cite{jochym2014using15}\cite{nikahd2016non16} and pieceable fault-tolerance \cite{yoder2016universal21} are well-known examples of such approaches.

Magic state distillation (MSD) is a procedure which uses only Clifford operations to increase the fidelity of non-stabilizer states that can be used to realize non-Clifford gates. This procedure is orders of magnitude more costly than transversal gates and can incur a significant resource overhead for the implementation of a quantum computer \cite{jochym2014using15}\cite{fowler2012surface19}.

The code switching method achieves a universal set of transversal gates by switching between two different quantum codes $C_1$ and $C_2$ where the non-transversal gates on $C_1/C_2$ have transversal implementations on $C_2/C_1$. To do so, a fault-tolerant switching network is required to convert $C_1$ into $C_2$ and vice versa. A general approach to convert between the codes is the use of teleportation \cite{choi2015dual14}\cite{oskin2002practical20}. Alternatively, a method has been proposed by Anderson et al. for direct fault-tolerant conversion between codes of Reed-Muller code family \cite{anderson2014fault10}. Moreover, a method has been recently published in \cite{yoder2016universal21} using pieceably fault-tolerant gates.

The code concatenation method uses two different quantum codes, namely $C_1$ and $C_2$ in concatenation to achieve universal fault tolerance. In this approach, the information is encoded by $C_1$ in the first level of concatenation and then all qubits of $C_1$ (uniform approach \cite{jochym2014using15}) or some of them (non-uniform approach \cite{nikahd2016non16}) are in turn encoded into the code of $C_2$. To this end, for any logical gate in the universal gate set with non-transversal implementation on $C_1$, there must exist an equivalent transversal implementation on $C_2$. Although this approach eliminates the need for magic state distillation, the number of necessary physical qubits to code the logical information is large. Moreover, for the non-transversal gates on $C_1$ as well as the non-transversal gates on $C_2$, the overall distance of the concatenated code is sacrificed.

Recently, Yoder et al. \cite{yoder2016universal21} proposed a novel approach to overcome the limits of non-transversality, namely pieceable fault-tolerance method (PFT). In this approach, a non-transversal circuit is broken into fault-tolerant pieces with rounds of intermediate error correction in between them to correct errors before they propagate to a set of non-correctable errors. As an example, fault-tolerant implementation of CCZ was developed for the 7-qubit Steane code. This considerably reduces the resource overhead in comparison with MSD. However, this approach is unable to find a fault-tolerant construction for non-transversal single-qubit gates, such as $T$, without additional ancillae. 

Despite substantial research on universal FTQC, the overhead of proposed approaches is still the main challenge for large-scale quantum computer design. In this paper, a hybrid approach is proposed which combines the code concatenation with code switching, PFT or MSD, to provide a low-overhead universal fault-tolerant scheme. 

\section{\label{sec:proposed}The proposed approach} \label{sec:proposed}
Similar to the code concatenation approach, the proposed method encodes the information using $C_1$ in the first level of concatenation and then the qubits of $C_1$ are in turn encoded into the code of $C_2$, either uniformly or non-uniformly. As there is no quantum code with a universal set of transversal gates, there is at least one non-transversal gate $G$ on $C_1$. Suppose that a circuit $U$ is the non-transversal implementation of $G$ on $C_1$ which is constructed using some gates $g_i$. In the proposed approach there may exist some gates $g_i$ with non-transversal implementation on $C_2$. This is in contrast to the code concatenation approaches where all of the $g_i$ gates must be transversal on $C_2$. Indeed, the proposed method uses more efficient code than code concatenation approaches in the second level of concatenation but with the overhead of using more costly approaches such as code switching, MSD or PFT for applying non-transversal gates. The idea behind this method is that the number of such non-transversal gates may be relatively small. 

Based on the implementation of the non-transversal gate $G$, the qubits of $C_1$ can be partitioned into two separate sets, namely $B_1$ and $B_2$. $B_1$ contains the coupled qubits and $B_2$ consists of the remaining qubits. In the proposed approach, the qubits of $B_1$ should be encoded using $C_2$ in the second level of concatenation whereas the qubits of $B_2$ can be left unencoded, encoded using $C_1$ or encoded using $C_2$. We refer to these three cases in dealing with the qubits of $B_2$ as Case 1, Case 2 and Case 3, respectively. The ways in which the gates are applied in the proposed approach are as follows.

The shared transversal gates on both $C_1$ and $C_2$ are globally transversal on the concatenated code and are therefore, fault-tolerant. For the transversal gates on $C_1$ with non-transversal implementation on $C_2$, although a single physical error may corrupt a $C_2$ logical qubit, it can be corrected using error correction procedure on $C_1$, similar to the code concatenation approach. 

The main challenge is fault-tolerant application of the non-transversal gates on $C_1$ referred to as $G$. As mentioned above, the implementation of $G$ on $C_1$ uses some gates $g_i$. The $g_i$ gates can be partitioned into two non-overlapping sets, namely $S_1$ and $S_2$. A gate $g_i$ belongs to $S_2$ if it has a transversal implementation on $C_2$ and belongs to $S_1$, otherwise. The gates of $S_2$ are transversal and therefore, fault-tolerant in the second level of concatenation. However, the proposed method exploits other existing approaches such as code switching, MSD or PFT for fault-tolerant application of the $S_1$ gates as they are non-transversal on $C_2$. Therefore, each $g_i$ gate is applied fault-tolerantly in the second level and a single error on one of the qubits of $B_1$ causes only a single physical error in each of the $B_1$ qubits which are themselves encoded blocks of $C_2$. Consequently, this error can be corrected using error correction procedure on $C_2$. 

This approach can lead to a low-overhead fault-tolerant implementation of the non-transversal gate $G$ if the number of non-transversal $S_1$ gates is relatively small for the selected code $C_1$. Fortunately, for a stabilizer code $[[n, 1, d]]$, a logical $C^{k}Z(\theta)$ gate can be implemented non-transversally using some Clifford gates and only one physical $C^{k}Z(\theta)$ gate. 
Therefore, for a non-transversal $C^{k}Z(\theta)$ on both $C_1$ and $C_2$, we have $|S_1|=1$ where the Steane code has been selected as $C_2$ as Clifford gates are transversal on Steane.

Let us now describe the proposed method in details by some examples using the 5-qubit perfect code, 7-qubit Steane code and 15-qubit Reed-Muller code (RM). Although the following examples are described based on the combination of code concatenation and code switching in two levels of concatenation, one can easily replace the code switching with MSD or PFT with no considerable modification and also generalize it for higher levels of concatenation, as explained later. It is reminded that PFT is unable to apply single-qubit gates such as $T$, fault-tolerantly.


\subsection{The proposed method based on the Steane code as $C_1$} \label{subsec:steaneCode}
\begin{figure}
	\centering
	\includegraphics[trim=0in 0in 0in 0in,clip,width=0.9\columnwidth]{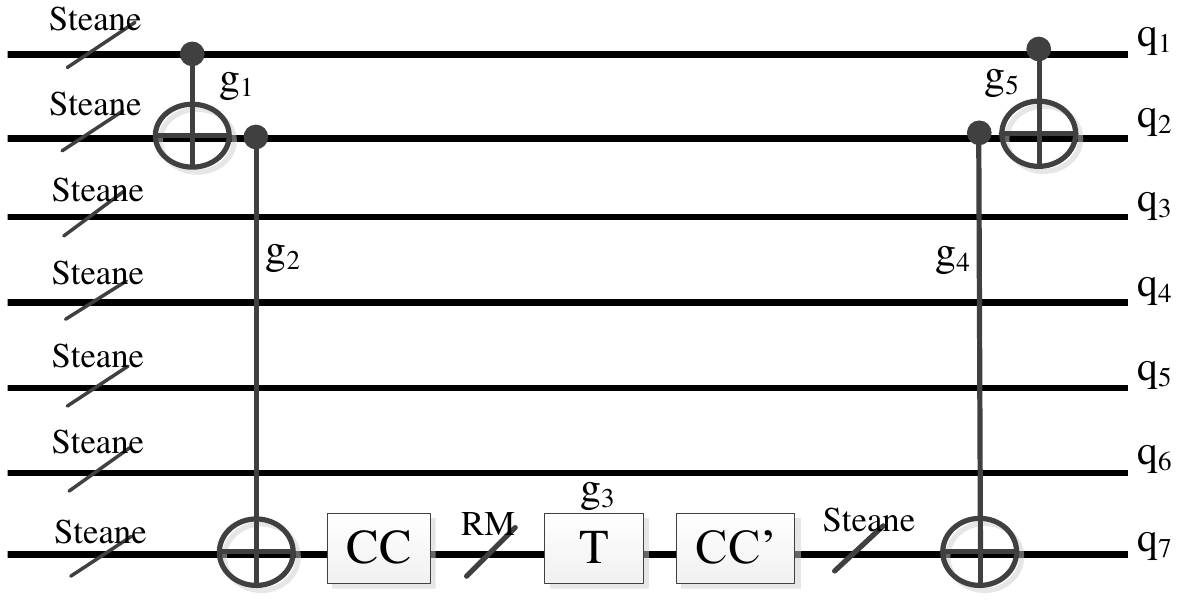}
	\caption{Fault-tolerant implementation of $T$ based on the proposed approach for the 49-qubit code.}
	\label{fig:steane_T}
\end{figure}

The Steane code is considered as $C_1$ in this section. The Clifford set \{$H$, $S=C^{0}Z(\frac{\pi}{2})$, $CZ=C^{1}Z(\pi)$\} along with a non-Clifford gate such as $T=C^{0}Z(\frac{\pi}{4})$ construct a universal set. Clifford gates are transversal on Steane while $T$ is not. The non-transversal implementation of $T$ on a Steane code block consists of one $T$ and four CNOT gates as shown in Fig. \ref{fig:steane_T}. By choosing the Steane code as $C_2$, we have $S_1=\{g_3\}$, $S_2=\{g_1, g_2, g_4, g_5\}$, $B_1=\{q_1, q_2,q_7\}$ and $B_2=\{q_3, q_4, q_5, q_6\}$. The qubits of $B_1$ are encoded using the Steane code and the qubits of $B_2$ can be encoded using Steane or left unencoded. Based on whether the $B_2$ qubits are left unencoded or are encoded using Steane, a 25- or 49-qubit code is produced, respectively.

Clifford gates are transversal in both levels of coding and are thus, fault-tolerant for the proposed concatenated codes. For fault-tolerant implementation of the logical $T$ gate, the gates of $S_2$ are applied transversally on Steane ($C_2$) and the $T$ gate ($g_3$) can be applied by switching $q_7$ from Steane into $RM$ and converting it back to Steane after transversal application of $T$ on $RM$.
Fig. \ref{fig:steane_T} shows this procedure for the 49-qubit code. $CC$ is an abbreviation for \emph{Code Converter} which can be implemented based on direct fault-tolerant conversion method between Reed-Muller codes \cite{anderson2014fault10}, efficiently. $CC$ and $CC$' convert the Steane code into the RM code and vice versa, respectively.

Note that this implementation of $T$ is not fully transversal and a single physical error on one of the $q_1$, $q_2$ or $q_7$ qubits can spread to another one. However, as the $CNOT$ and $T$ gates are implemented transversally in the second level of concatenation, this error propagates to at most one physical qubit in each code block which can be corrected using error correction procedure on the code blocks in the second level of concatenation. Therefore, any arbitrary single physical error can be corrected during the application of $T$ gate.

The $CCZ=C^{2}Z(\pi)$ gate can also be applied fault-tolerantly for the proposed 25- and 49-qubit codes, as its implementation on the Steane code has the same structure as $T$ and it is transversal on $RM$.

\subsection{The proposed method based on the 5-qubit code as $C_1$} \label{subsec:5qubitCode}
\begin{figure}
	\centering
	\includegraphics[trim=0in 0in 0in 0in,clip,width=0.9\columnwidth]{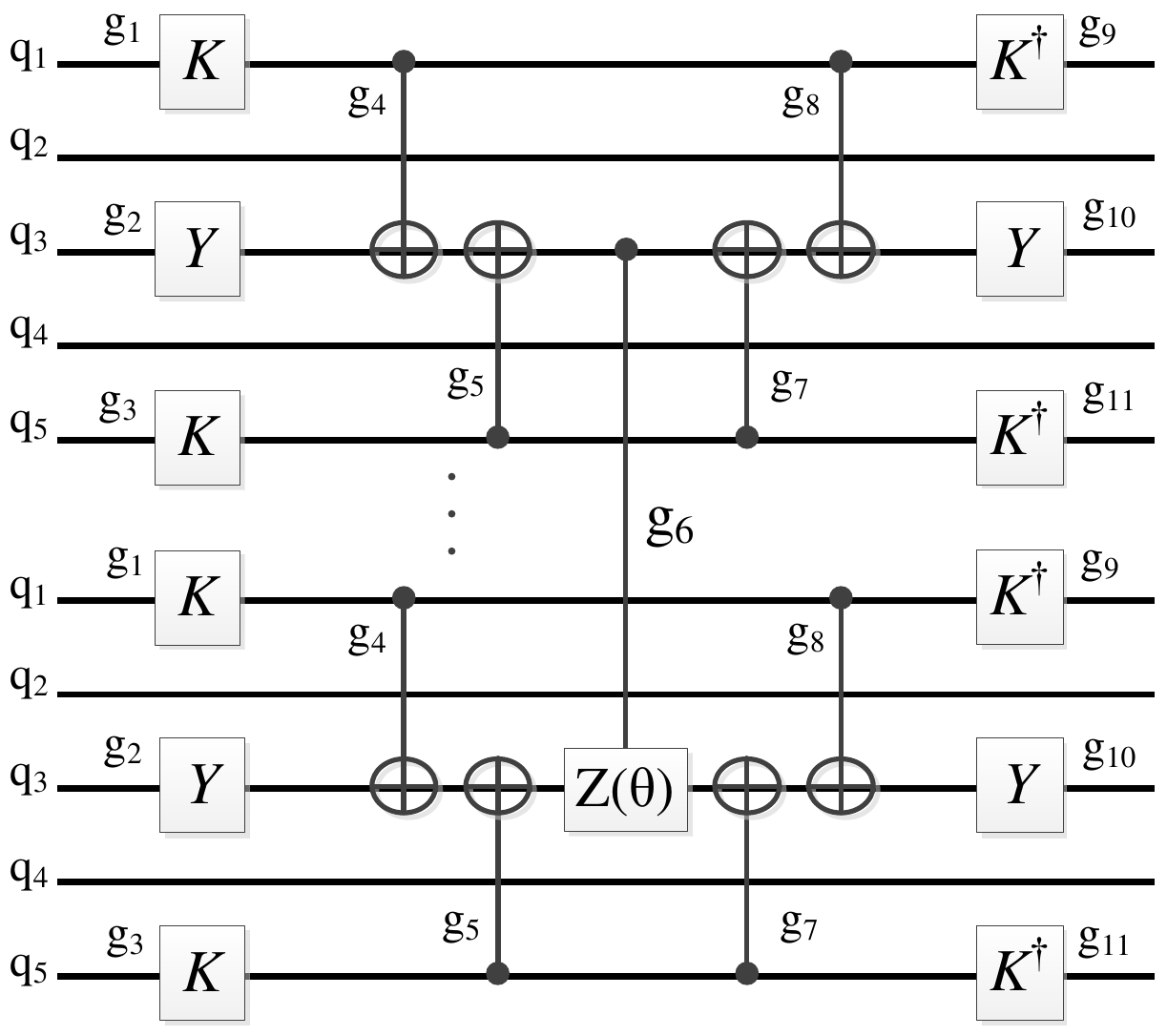}
	\caption{Non-transversal implementation of the $C^{k}Z(\theta)$ gate for the 5-qubit code.}
	\label{fig:CKZ5}
\end{figure}

In this section, the 5-qubit code is selected as $C_1$ and a logical qubit is encoded into the 5-qubit code in the first level of concatenation. Let M=\{$T=C^{0}Z(\frac{\pi}{4})$, $S=C^{0}Z(\frac{\pi}{2})$, $CZ=C^{1}Z(\pi)$, $CCZ=C^{2}Z(\pi)$\}. The gates in $M$ along with $K$ form a universal set for quantum computation, where $K=SH$. The $K$ gate is transversal on the 5-qubit code, however, the gates of $M$ are not. The gates of $M$ belong to the class of $C^{k}Z(\theta)$ gates and thus, as described before, Steane can be selected as $C_2$. Based on Fig. \ref{fig:CKZ5}, that shows the non-transversal implementation of $M$ gates on the 5-qubit code, we have $B_1=\{q_1, q_3, q_5\}$ and $B_2=\{q_2, q_4\}$ and also $S_1=\{g_6\}$ and $S_2=\{g_1, g_2, g_3, g_4, g_5, g_7, g_8, g_9, g_{10}, g_{11}\}$ only for the $T$ and $CCZ$ gates (note that $S$ and $CZ$ are transversal on Steane). The qubits of $B_1$ are encoded using Steane and the qubits of $B_2$ can be either left unencoded or encoded using the 5-qubit code or Steane which leads to a 23-, 31- or 35-qubit code, respectively.

The $K$ gate can be applied transversally for all of the 23-, 31- and 35-qubit codes, as it is transversal on both the 5-qubit and Steane codes. The $S$ and $CZ$ gates are transversal on Steane. Consequently, these gates can be applied fault-tolerantly on the concatenated code without need to exploit code switching. However, for fault-tolerant implementation of $T$ and $CCZ$, the proposed method dynamically alternates between Steane and RM for $q_3$ when the $T$ or $CCZ$ gate ($g_6$) should be applied. 

\section{Code analysis} \label{sec:codeAnalysis}
Straight concatenation of the two codes [[$n_1$,1,$d_1$]] and [[$n_2$,1,$d_2$]] leads to a code [[$n_1$$n_2$,1,$d_1d_2$]] \cite{gottesman1997stabilizer22}. However, this distance ($d_1d_2$) may be sacrificed because of error propagation during application of the non-transversal gates. We refer to $d_1d_2$ as the overall distance of the code and use \emph{effective distance} to specify the sacrificed distance. In this section, we analyze the proposed codes in terms of overall and effective distance. 

Let $d_1$ and $d_2$ be the distance of $C_1$ and $C_2$, respectively. For the concatenated codes with fully encoded qubits in both levels of concatenation (Cases 2 and 3), the overall distance will be $d_1d_2$ while for the codes with unencoded $B_2$ qubits (Case 1), the overall distance is reduced. For example, the proposed 31-, 35- and 49-qubit codes have the overall distance of 9 and the overall distance of the 23- and 25-qubit codes are 5 \cite{chamberland2016architectural23}.

The effective distance of the proposed codes varies for different gates. Table \ref{tab:comparison} compares the effective distance of the proposed codes for the gates of the selected universal sets. 

\begin{table}%
	\begin{center}
		\caption{Comparison of the proposed concatenated codes in terms of the number of qubits and effective distance for different gates.}
		{\footnotesize
			\begin{tabular}{ |c|c|c|c|c|c|c|c|c|c| }
				\hline
				$C_1$ & Case & \# qubits & $H$ & $K$ & $T$ & $S$ & $CZ$ & $CCZ$ & worst case \\ \hline
				\multirow{2}{*}{Steane} & 1 & 25 & 5 & 5 & 3 & 5 & 5 & 3 & 3\\ \cline{2-10}
				& $2 \equiv 3$ & 49 & 9 & 9 & 3 & 9 & 9 & 3 & 3\\ \hline
				\multirow{3}{*}{5-qubit} & 1 & 23 & - & 5 & 3 & 3 & 3 & 3 & 3\\ \cline{2-10}
				& 2 & 31 & - & 9 & 3 & 3 & 3 & 3 & 3\\ \cline{2-10}
				& 3 & 35 & 9 & 9 & 3 & 3 & 3 & 3 & 3\\ \hline
			\end{tabular}
		\label{tab:comparison}
		}
	\end{center}
\end{table}

For the shared transversal gates on both $C_1$ and $C_2$, no error is propagated in the code blocks and thus, the effective distance of the concatenated codes will be equal to its overall distance. The gates with effective distance of 5 and 9 in Table \ref{tab:comparison} are examples of such gates. Note that the $H$ gate is applicable for the 35-qubit code with the effective distance of 9. This is because $H$ is transversal on the 5-qubit code by permutation \cite{yoder2016universal21}. This permutation is applicable for the 35-qubit code, as all qubits in the first level are encoded using Steane in the second level of concatenation. 
However, application of $H$ with permutation for the 23 and 31-qubit codes destroys the code structures as they have code blocks encoded using different codes in the second level. Generally, a transversal gate with permutation on $C_1$ is not applicable for non-uniform concatenated codes \cite{nikahd2016non16}, unless it permutes only the encoded blocks of the same code.


Generally, for the non-transversal gates on $C_1$, the effective distance is different for Case 1, Case 2 and Case 3. For Case 1, the effective distance is $min(d_1, d_2)$. 
In this case, the qubits of $B_1$ are encoded blocks of $C_2$ and the qubits of $B_2$ are left unencoded. The errors may propagate between the qubits of $B_1$ and therefore, in the worst case either $t_1$ errors on $B_2$ or $t_2$ errors on $B_1$ can be corrected, where $t_1=\lfloor \frac{d_1-1}{2}\rfloor$ and $t_2=\lfloor \frac{d_2-1}{2}\rfloor$. For Case 2, the qubits of $B_1$ and $B_2$ are encoded blocks of $C_2$ and $C_1$, respectively. In the worst case, either $\lfloor \frac{d_1^{2}-1}{2}\rfloor$ errors on $B_2$ or $t_2$ errors on $B_1$ can be corrected and therefore, the effective distance of the code will be $min(d_1^{2},d_2)$. In Case 3, where the qubits of $B_2$ are also encoded blocks of $C_2$ the effective distance is $min(d_1d_2, d_2) = d_2$, as $t_2$ errors on qubits of $C_1$ propagate to at most $t_2$ errors on each qubit of $B_1$ and these errors can be corrected using error correction procedure on $C_2$ code blocks. \{$S$, $CZ$, $T$, $CCZ$\} and \{$T$, $CCZ$\} are such gates for the proposed codes based on the 5-qubit and Steane codes, respectively. 
 
In general, for the gates that are transversal on $C_1$ but non-transversal on $C_2$, the effective distance is $d_1$. This is because in the worst case, although $t_1$ errors on $t_1$ distinct qubits of $C_1$ may lead to $t_1$ logical errors on the second level, these errors can be corrected using error correction procedure on $C_1$. However, there exist no such gates in the code examples in this paper.

The proposed code examples outperform the codes based on code concatenation proposed in \cite{jochym2014using15} and \cite{nikahd2016non16} as they need fewer qubits and less resource to protect the computation from arbitrary single physical error. 
Furthermore, the overall distance of the proposed codes is compromised for only the non-transversal gates on $C_1$ as there is no gate with transversal implementation on $C_1$ and non-transversal implementation on $C_2$ whereas in the previous code concatenation approaches, the overall distance of the code is sacrificed for the non-transversal gates on both $C_1$ and $C_2$. For example, the counterparts of the proposed 25- and 49-qubit codes have 49 \cite{nikahd2016non16} and 105 qubits \cite{jochym2014using15}, respectively. Moreover, for the $H$ gate, the proposed 25- and 49-qubit codes have the effective distance of 9 while the effective distance of their concatenated counterparts, e.g. 49- and 105-qubit codes, have been reduced to 3. This result becomes more valuable by the fact that the threshold of the 49- and 105-qubit concatenated codes are limited by the application of logical $H$ gate \cite{chamberland2016thresholds22}\cite{chamberland2016architectural23}. The only overhead of the proposed codes in comparison with concatenated codes is using code switching, MSD or PFT for application of the $S_1$ gates (e.g. $T$ or $CCZ$).

In comparison with the code switching, MSD and PFT approaches, the proposed method significantly reduces the implementation overhead of non-transversal gates in two-level concatenated codes. The main disadvantage of our method is that the overall distance of the concatenated code is sacrificed for the non-transversal gates in comparison with 
these approaches. 

%

\section{\label{sec:discussion}Discussion}


One may use MSD or PFT instead of code switching to achieve FTQC. In this case, the gates $g_i \in S_1$ are applied based on these methods in the second level of concatenation. For example, for the Steane code, the $T$ gate ($g_3$) can be easily applied based on MSD instead of code switching. Generally, each gate $g_i \in S_1$ can be implemented separately using appropriated method according to the resulting cost.

The proposed method can be extended to an $l$-level concatenated code in a way that the non-transversal gate $G$ is implemented based on its non-transversal implementation on $C_1$ and code switching (MSD or PFT) for the first ($l-k$) and the last $k$ levels of concatenation, respectively. This results in a trade-off between the effective distance of the code and the resource overhead for applying the non-transversal gate $G$. The $k$ value may change between 1 and $l$. For $k=1$, the proposed method leads to the lowest overhead for implementing $G$ and for $k=l$, the method will be the same as code switching (MSD or PFT) where the overall distance of the concatenated code is achievable but at the cost of highest overhead.
a
It is worth mentioning that there is no need for considerable additional ancillary qubits and resources for the implementation of non-transversal gates. Therefore, we can eliminate the need for consideration of a special area in the architecture, dedicated to the execution of these gates. 

\section{\label{sec:conclusion}Conclusion}

In this article, we proposed a low-overhead hybrid approach for universal FTQC. The proposed method combines code concatenation with code switching, MSD or PFT schemes. This method outperforms code concatenation approaches for a code $C_1$ with small sizes of $S_1$. Particularly, for a stabilizer code $[[n, 1, d]]$, a $C^{k}Z(\theta)$ gate has an efficient non-transversal implementation with $|S_1|=1$. 

The proposed approach was described based on the 5-qubit and Steane codes in two levels of concatenation as examples which leads to the 25-, 49-, 23-, 31- and 35-qubit codes. The proposed codes have significantly fewer number of qubits in comparison with the codes based on code concatenation approaches. In addition, the effective distance of the proposed codes are only compromised for non-transversal gates on $C_1$ as there is no transversal gate on $C_1$ with non-transversal implementation on $C_2$. Furthermore, this approach significantly reduces the resource overhead in comparison with code switching, MSD and PFT at the cost of reducing the effective distance of the concatenated code for implementing non-transversal gates.

Making an accurate estimation of the error threshold and resource overhead of the proposed codes and exploring the method for other codes have been considered as future works.

\bibliography{NSS}

\end{document}